\documentclass{aastex}
\usepackage{emulateapj5,apjfonts,epsfig}

\def\asca{{\sl ASCA }}

\def\xte{{\sl RXTE }}

\def\chandra{{\sl Chandra }}
\def\ergsec{\hbox{erg s$^{-1}$ }}
\def\ergcm{\hbox{erg cm$^{-2}$ s$^{-1}$ }}

\def\it{\sl}

\def\lapp{\ifmmode\stackrel{<}{_{\sim}}\else$\stackrel{<}{_{\sim}}$\fi}

\def\gapp{\ifmmode\stackrel{>}{_{\sim}}\else$\stackrel{>}{_{\sim}}$\fi}
\def\spose#1{\hbox to 0pt{#1\hss}}

\def\approxlt{\mathrel{\spose{\lower 3pt\hbox{$\sim$}}
        \raise 2.0pt\hbox{$<$}}}
\def\approxgt{\mathrel{\spose{\lower 3pt\hbox{$\sim$}}
        \raise 2.0pt\hbox{$>$}}}

\slugcomment{Submitted for publication to The Astrophysical Journal}

\shorttitle{X-RAY P~CYGNI LINE VARIABILITY FROM CIR~X-1}

\shortauthors{SCHULZ \& BRANDT}

\begin{document}

\title{Variability of the X-ray P~Cygni Line Profiles from Circinus~X-1 Near Zero Phase}
\author{
N. S. Schulz\altaffilmark{1}
and
W. N. Brandt\altaffilmark{2}}
\altaffiltext{1}{Center for Space Research, Massachusetts Institute of Technology,
Cambridge, MA 02139.}
\altaffiltext{2}{Department of Astronomy \& Astrophysics, 525 Davey Laboratory,
The Pennsylvania State University, University Park, PA, 16802.}

\begin{abstract}
The luminous X-ray binary Circinus~X-1 has been observed twice near zero orbital phase using
the High-Energy Transmission Grating Spectrometer (HETGS) onboard {\sl Chandra}. The source 
was in a high-flux state during a flare for the first observation, and it was in a low-flux 
state during a dip for the second. Spectra from both flux states show clear P~Cygni lines,
predominantly from H-like and He-like ion species. These indicate the presence of a 
high-velocity outflow from the Cir~X-1 system which we interpret as an equatorial 
accretion-disk wind, and from the blueshifted resonance absorption lines we determine outflow 
velocities of 200--1900~km~s$^{-1}$ with no clear velocity differences between the two flux 
states. The line strengths and profiles, however, are strongly variable both between the two 
observations as well as within the individual observations. We characterize this variability 
and suggest that it is due to both changes in the amount of absorbing material along the 
line of sight as well as changes in the ionization level of the wind. We also 
refine constraints on the accretion-disk wind model using improved plasma 
diagnostics such as the He-like Mg~XI triplet, and we consider the possibility that 
the X-ray absorption features seen from superluminal jet sources can generally be 
explained via high-velocity outflows. 
\end{abstract}

\keywords{
stars: individual (Cir~X-1) ---
stars: neutron ---
X-rays: stars ---
binaries: close ---
accretion: accretion disks ---
techniques: spectroscopic}

\section{Introduction}

The discovery of broad P~Cygni line profiles from the luminous X-ray
binary Circinus~X-1 (hereafter Cir~X-1), which we reported in
Brandt \& Schulz (2000; hereafter Paper~I), was the first time such lines
had been observed clearly in the X-ray spectrum of a cosmic object. The
P~Cygni lines demonstrated the presence of a high-velocity outflow, in
line with other suggestions of outflow in this system
(e.g., Johnston, Fender, \& Wu 1999). However, the nature of
Cir~X-1 in general is still poorly understood and, despite advances
in recent years, there remains uncertainty about even the most
basic properties of this system. Since its discovery (Margon et~al. 1971),
it has appeared bright and variable in X-rays exhibiting a period of 16.6 days
(Kaluzienski et~al. 1976), and it has thus been a frequent target of most
X-ray observatories. The compact object in the Cir~X-1 system is thought
to be a neutron star (Tennant, Fabian, \& Shafer 1986) that can radiate
at super-Eddington luminosities during times of strong mass transfer. Its
heavily reddened optical counterpart (e.g., Moneti 1992) shows strong,
asymmetric H$\alpha$ emission; this emission is variable and appears to
arise from several sites in the system including the accretion
disk (e.g., Whelan et~al. 1977; Mignani, Caraveo, \& Bignami 1997;
Johnston et~al. 2001). The system shows two arcminute-scale
radio jets (Stewart et~al. 1993), and an arcsecond-scale asymmetric
jet (Fender et~al. 1998) suggests the presence of relativistic outflow
from the source. Cir~X-1 is often included among the ``Galactic
microquasar'' X-ray binaries (Mirabel 2001). 
  
The nature of the companion star is an important issue for the
interpretation of the observed X-ray P~Cygni lines. Specifically,
one must address if the P~Cygni lines could be made by absorption
of X-rays from the neutron star in a high-velocity wind from the
companion; such a scenario is likely for the high-mass X-ray binary
Cygnus~X-3, which also shows X-ray P~Cygni lines (Liedahl et~al. 2000)
and has a Wolf-Rayet star companion. The large velocities observed
for the X-ray P~Cygni lines of Cir~X-1 (up to $\pm 1900$~km~s$^{-1}$)
would require a high-mass companion, probably of spectral type~O.

Several attempts have been made to determine the spectral type of the
companion star. While a low-mass companion is favored by most
of the recent studies (e.g., Stewart et~al. 1991; Glass 1994), the
constraints remain weak due largely to the heavy interstellar reddening.
Additional evidence for a low-mass companion comes from observations
of the correlated X-ray spectral and timing properties, observations
of type~I X-ray bursts, and theoretical considerations. For example,
Shirey, Bradt, \& Levine (1999a) and Shirey, Levine, \& Bradt (1999b) have demonstrated
in extensive studies of \xte data that Cir~X-1 exhibits spectral branches in
the hardness-intensity diagram that can be identified with the horizontal,
normal, and flaring branches of ``Z'' type low-mass X-ray binaries (LMXBs;
e.g., Hasinger \& van der Klis 1989; Schulz, Hasinger, \& Tr\"umper 1989).
Qu, Yu, \& Li (2001) have also noted some similarities to the LMXBs GX~$5-1$
and Cyg~X-2. Furthermore, Tauris et~al. (1999) considered the (somewhat
uncertain) kinematic properties of the Cir~X-1 system and argued that
the companion is a low-mass ($\lesssim 2.0$~$M_\odot$), unevolved star.

Given that the stellar wind from the likely low-mass companion cannot
explain the observed properties of the X-ray P~Cygni profiles from Cir~X-1,
we interpreted these profiles in the context of an equatorial wind driven
from the accretion disk by a combination of Compton heating and radiation
pressure (see Paper~I). The fact that Cir~X-1 can radiate with high
luminosity relative to its Eddington luminosity ($L/L_{\rm Edd}$) makes
it a natural system in which to expect observable outflows, because
of the larger amount of photon pressure available per unit gravitational
mass. Accretion-disk winds in X-ray binaries have been discussed both
theoretically and observationally (e.g., Begelman, McKee, \& Shields 1983;
Raymond 1993; Chiang 2001; Proga \& Kallman 2001), and the P~Cygni lines
seen from Cir~X-1 match the lines predicted by Raymond (1993) reasonably
well. We also note that Iaria et~al. (2001a, 2001b) have recently found
evidence for a large column density of ionized gas along the line of
sight based on observations of ionized iron~K edges; the same gas might
well produce both the X-ray P~Cygni lines and the ionized iron~K edges.

P~Cygni lines are a common property of the ultraviolet spectra of
cataclysmic variables (CVs) possessing an accretion disk. In these
systems, the line profiles are associated with high-velocity
polar outflows (e.g., C\'ordova \& Howarth 1987), in contrast
to the equatorial geometry proposed for the accretion-disk wind 
in Cir~X-1. The proposed equatorial geometry was 
motivated by evidence that the Cir~X-1 system is viewed in a relatively
edge-on manner. Although eclipses are not observed, the spectral
variability caused by the observed X-ray absorption is best explained by
a model with a relatively edge-on accretion disk (Brandt et~al. 1996;
Shirey et~al. 1999b). In this model, X-rays created near
the neutron star reach the observer via two light paths: (1) a direct,
but often absorbed, light path that can intersect the outer bulge of the
accretion disk, and (2) an indirect, electron-scattered light path
that avoids the absorption associated with the direct light path.
This model can naturally explain the large observed changes in
the column density of the absorbing gas without corresponding
changes in its apparent covering fraction (see Brandt et~al. 1996
for details). We note that radio imaging of SS433, which shares
several similarities with Cir~X-1, provides strong evidence for
an equatorial wind-like outflow from this system (Blundell et~al. 2001). 

In this second paper, we present results from a variability study
of the X-ray P~Cygni lines from Cir~X-1. As in Paper~I the
observations were performed with the High-Energy Transmission Grating
Spectrometer (HETGS) onboard \chandra at ``zero phase.'' Zero phase in
Cir~X-1 is thought to be associated with the periastron passage of
the neutron star, and near this phase Cir~X-1 is highly variable in
X-ray intensity. We present analyses of two observations, which
include a re-assessment of the data presented in Paper~I representing
a high-flux state of Cir~X-1 and a second observation during
which Cir~X-1 was in a low-flux state with strong variability.
In Schulz \& Brandt (2001; hereafter SB2001) we already presented a
broad-band continuum fit to some of the data as well as showed temporal
changes of the Si~XIV line. The analysis here focuses on a comparison
of the bright lines in the two zero-phase flux states and
their variability with intensity and time.

Throughout this paper, we adopt a distance to Cir~X-1 of 6~kpc
(Stewart et~al. 1993; Case \& Bhattacharya 1998) and an interstellar 
column density of $2\times10^{22}$ cm$^{-2}$ (e.g., Predehl \& Schmitt 1995). 

\centerline{\epsfxsize=8.5cm\epsfbox{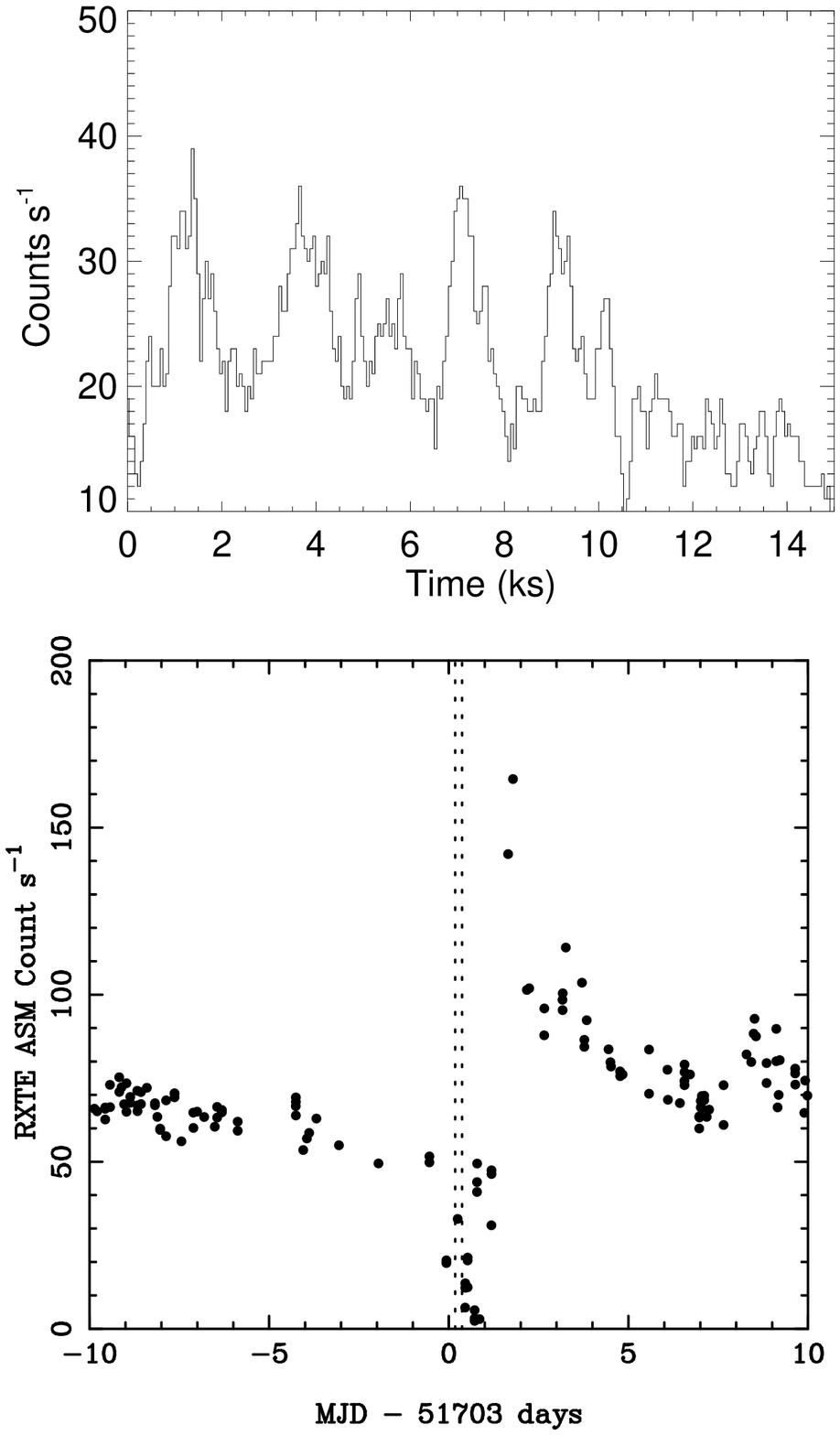}}
\figcaption{
The top panel shows the 1st order light curve of observation~II in 60~s
bins. The bottom panel shows the \xte ASM light curve including 10 days before
and after observation~II was performed with \chandra (the \chandra
observation window is marked by dotted lines). A similar plot for
observation~I can be found in Paper~I.\label{lightcurve}}

\section{Chandra Observations}

Cir~X-1 was observed with the \chandra HETGS (C.R. Canizares et~al., in preparation) 
on 2000 February~29 (starting at 22:09:50 UT) for 30~ks (observation~I) and on 
2000 June~8 (starting at 04:13:01 UT) for another 15~ks 
(observation~II).\footnote{For details on \chandra and the HETGS see
the \chandra Proposers' Observatory Guide at http://asc.harvard.edu/udocs/docs/.} 
Both observations were performed near zero phase of the binary orbit. 
They were not precisely at the same phase; observation~I started at phase
0.9917, and observation~II started at phase 0.9923 (adopting the ephemeris
of Glass 1994). In Paper~I we showed that observation~I
occurred during the onset of a luminous flare. Figure~1 puts observation~II
similarily into perspective with the long-term \xte ASM light curve. Here it
appears that we observed the source during a prominent intensity dip. 
Observation~I is also different from observation~II because 
it suffered from a misalignment of the applied subarray during the observation
which led to the loss of half of the data. Otherwise observation~II was
performed with a similar instrumental setup; the zeroth order
was blocked from being telemetered, and the same subarray with CCDs S0 and S5
inactive was applied leading to a frame time of 1.7~s.    

\centerline{\epsfxsize=8.2cm\epsfbox{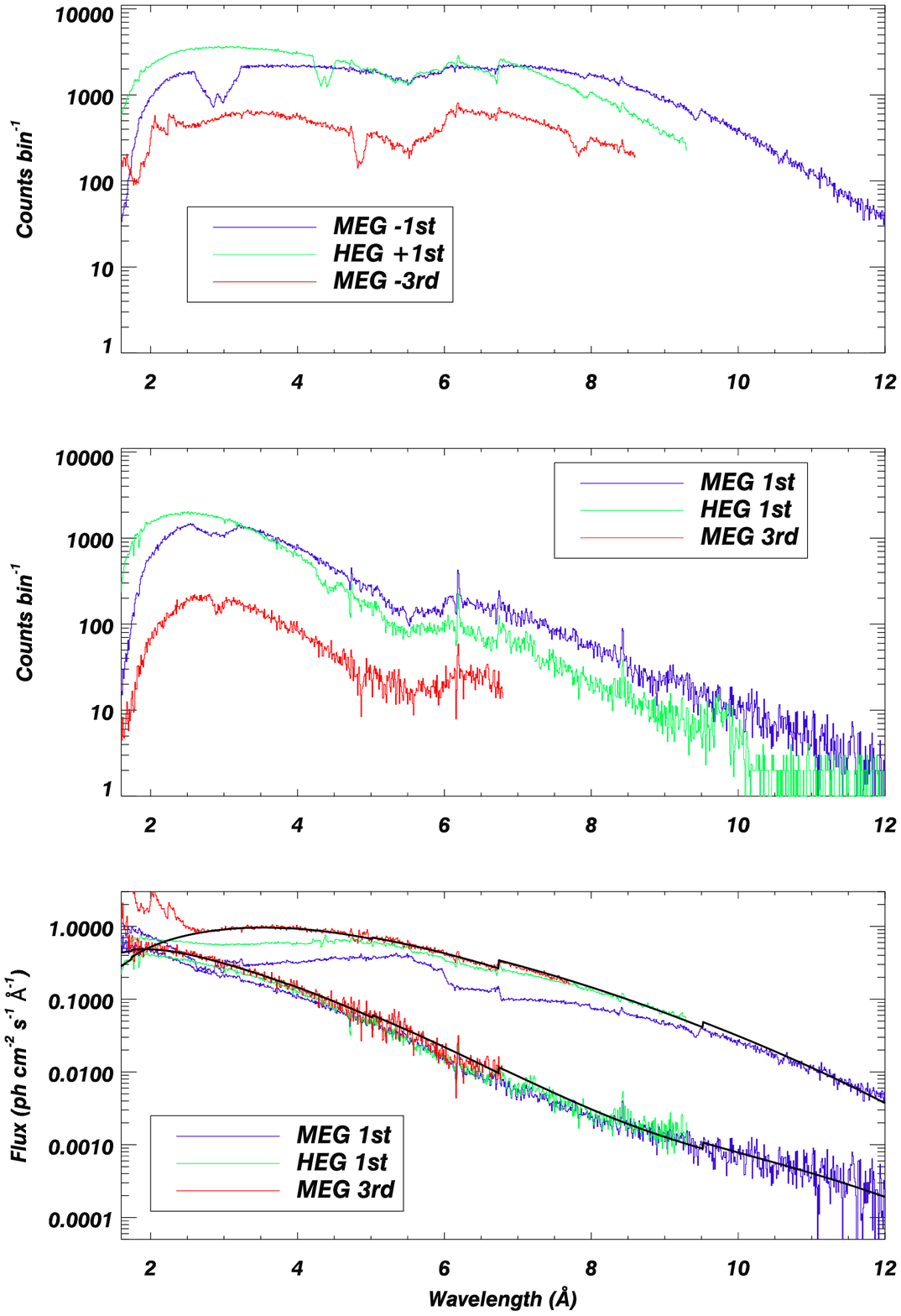}}
\figcaption{The raw-count spectra of observation~I (top) and observation~II (middle) plotted
on the same scale. The MEG 1st order spectra are blue, the HEG 1st order spectra are green,
and the MEG 3rd order spectra are red. The bottom panel shows these corrected for exposure.
The brighter set corresponds to observation~I, and the fainter set corresponds to observation~II.
\label{counts}}

\subsection{Raw-count spectra}

Observation~I was the basis for the discovery of the P~Cygni profiles, and 
most of its basic properties were described in Paper~I.  
We reprocessed these data using several
improved calibration data products. The first one concerns a small correction
to the wavelength scale, which arose from a re-assement of the detector
pixel size and which led to a scale factor of 500 ppm in the wavelength
scale. This leads to a maximal difference of 0.0075~\AA\ at 15~\AA\ compared
to the previous definition. For shorter wavelengths this change is of the
order of the previous accuracy of the wavelength determination. A second
change came from the application of the CCD quantum efficiency, which has now
been provided with a finer energy grid. The benefits here are that instrumental
edges, for our interest especially the CCD Si-edge, have been accounted
for more accurately and are less likely to produce spectral artifacts.
A third improvement came from correction of a slight mismatch of the 
energy grids in the ancillary data files and the grating response.
All data were processed the same way with the exception that in
observation~II we additionally added positive and negative orders together. In this
respect we re-gained essentially the same exposure as in observation~I.
The raw-count spectra in Figure~2 (top) indicate a very high number of 
counts during observation~I. The MEG 1st order spectra have $1.4\times 10^6$ events, 
the HEG 1st order spectra have $1.5\times 10^6$ events, and the MEG 3rd order 
spectra have $3.9\times 10^5$ events. 
  
In observation~II (middle) the MEG 1st order spectra have $3.3\times 10^5$ events,
the HEG 1st order spectra have $2.1\times 10^5$ events, and the MEG 3rd order 
spectra have $4.0\times 10^4$ events. The difference in intensity is well
illustrated by the raw-count spectra shown in Figure~2.
Throughout most of the spectral bandpass, the number of counts 
in observation~II is significantly lower in each bin compared
to observation~I. This had the favorable effects that we did not suffer from any 
frame dropouts during data transmission and were much less affected by pile-up. 
The number of counts per bin in observation~II drops below $\approx 1$ above 12~\AA,
which basically reflects the noise level. In Figure~2 we therefore terminate the
spectra at 12~\AA\ for better visibility of the details at shorter wavelengths, 
although the MEG spectrum in observation~I continues up to 15.5~\AA\ (see Paper~I). 

\subsection{Pile-up issues}

The amount of pile-up in the spectra makes detailed modeling of the 
X-ray continuum difficult. Pile-up affects the MEG and HEG spectra 
over a considerable bandpass where the number of incident counts per frame in a given 
pixel is very high. In order to get a lower limit of the number of detected
counts per frame and pixel a rough estimate of its dependence on wavelength,
one can divide the count spectra in Figure~2 by the number of CCD frames obtained during
the observation period and the number of pixels included in each bin. 
Values significantly above 0.01 indicate that the spectra are likely
affected by pile-up. In general to obtain more precise information 
there is no straightforward procedure, but there are several other 
leads that allow one to identify pile-up affected regions in the spectra.  
High photon densities are only reached in the 1st orders where 
the effective area of the instrument is highest. In the MEG 1st order 
spectrum this happens below $\sim$ 11~\AA, while in the HEG 1st order 
spectrum it happens below $\sim$ 8~\AA. All higher orders are not affected 
by pile-up. However, they still suffer from contamination
of piled-up photons from the 1st orders. Pile-up in the 1st orders creates a
depletion of photons at a specific wavelength. Since it is a CCD effect,
multiple photon coincidences appear in higher CCD channels since they are 
recorded as an event that has the sum of the incident energies.
In the HETGS these channels also coincide with the locations of the 
higher order grating spectra. In this respect three first-order photons, for 
example, appear as one photon in 3rd order at 1/3 of the original wavelength. 
This means that the MEG 3rd order below $\sim 2$~\AA\ is contaminated 
by pile-up from the MEG 1st order. Since no viable model yet exists to 
correct for this effect, we will avoid all areas in the spectra that 
suffer substantially from pile-up. In the MEG 1st order the range below 
2~\AA\ is also affected by some contamination from zero-order scattering
and is excluded as well. For observation~I, for example, this leaves the 
following wavelength bands for spectral modeling: 2.4--8.5~\AA\ in the 
MEG 3rd order, 8.2--10.1~\AA\ in the HEG 1st order, and above 11~\AA\ in 
the MEG 1st order. The wavelength range below 2~\AA\ in the HEG 1st order 
is mildly affected by pile-up, and here we have to add corrections (see \S3.5). 
The available energy bands are thus 0.8--5.2~keV and 6.2--8~keV. Observation~II shows
considerably lower count rates, and its spectra are much less affected by pile-up
and zero-order scattering; here we can use almost the entire 1.0--8.0~keV 
energy band. 

Figure~2 (bottom) shows the exposure-corrected spectra for observations~I and II.
The top set of spectra with the higher fluxes are from observation~I, and the 
bottom set are from observation~II.
The spectra are integrated over the entire exposures.
Without pile-up the blue, green and red
spectra of both sets should overlap. This
qualitatively gives an impression of the degree of pile-up, which is indicated by how much the
blue and green spectra deviate from the red spectra.

For the analysis of the lines we mostly rely on the HEG 1st order 
spectra for statistical reasons. This requires some corrections in areas 
where pile-up appears strong, which is between 2 and 7~\AA. From
the difference of the exposure-corrected spectra and the pile-up free 3rd order spectra we can
determine an energy dependent flux correction factor, which we later use to correct the
equivalent widths of the lines. The model of the HEG 1st order spectrum of observation~II 
in Figure~3 has these factors applied, and the figure shows that we get good agreement.
The additional systematic uncertainty in the determination of 
equivalent widths can be up to 20$\%$ is the worst case.

\subsection{Determining fluxes and luminosities}

In general we follow the procedure to determine the luminosity outlined in SB2001 and 
choose our model spectrum from previous fits to \asca data (Brandt et~al. 1996); the
model consisted of two blackbody components with both intrinsic partial-covering absorption
and interstellar absorption of $2.0\times 10^{22}$~cm$^{-2}$ using cross-sections 
from Morrison \& McCammon (1983). The two observations can be compared to the 
low-count-rate state (observation~II) and high-count-rate state (observation~I) in 
Brandt et~al. (1996). We emphasize, however, that the current HETGS spectra do not 
qualify for a detailed physical discussion of the continuum, and we use the fits 
mainly as a parameterization of the continuum emission in order to perform the line 
analysis. We also draw some basic information from the change of the fits from 
observation~I to observation~II.

For observation~I we find a soft blackbody temperature of $kT=0.66\pm 0.17$~keV 
[partial covering $N_{\rm H}=(8.2\pm 0.4)\times 10^{23}$~cm$^{-2}$] and a 
hard blackbody temperature of $kT=1.22\pm 0.28$~keV [partial covering 
$N_{\rm H}=(2.3\pm 0.2)\times 10^{22}$~cm$^{-2}$]; the partial covering fraction was 45$\%$.
We find a total absorbed X-ray flux of $2.0\times10^{-8}$~\ergcm with the soft component 
contributing 61$\%$ between 2--8~keV. This corresponds to a luminosity of 
$1.8\times 10^{38}$~erg~s$^{-1}$. During observation~II we could fix the
the partial covering columns to the values found in observation~I; we find 
a covering fraction of 89$\%$. The temperature for the soft blackbody component 
was $kT=0.16\pm 0.09$~keV, and for the hard blackbody component it was
$kT=1.55\pm 0.21$~keV. The total absorbed X-ray flux dropped to $4.8\times10^{-9}$~\ergcm 
with a contribution of the soft component of $<1$\%. The corresponding luminosity 
is $2.8\times 10^{37}$~\ergsec. We emphasize again that the model parameters we 
have empirically derived to parameterize the continuum are not optimal for
comparison with previous results (Brandt et al. 1996; Shirey et al. 1999).
A full pile-up model will be required for detailed HETGS continuum fitting,
and even then the limited high-energy sensitivity will remain an obstacle. As we 
discuss in \S3.5, the correction for absorption during observation~II
is generally uncertain, and hence the true luminosity of Cir~X-1 is difficult to constrain
during this observation. It is plausible that the true luminosity during observation~II 
is comparable to that during observation~I. Such behavior would be consistent with that 
observed by \asca (Brandt et~al. 1996) and \xte (Shirey et~al. 1999b) during dips. 
Evidence for heavy and complex absorption in the \chandra data is found in the
depths of the neutral Fe~K (see \S3.5) and maybe the Si~K edges. Both observations show a 
neutral Si~K edge with a depth of $\tau=0.15$. The large depth of this edge required 
us to add an additional edge to our fits. However, although current uncertainties in the improved instrument 
efficiencies cannot account for such a large effect, it was also observed recently in HETGS spectra
of other bright sources (Miller et al. 2002).

\begin{figure*}
\centerline{\epsfxsize=16.5cm\epsfbox{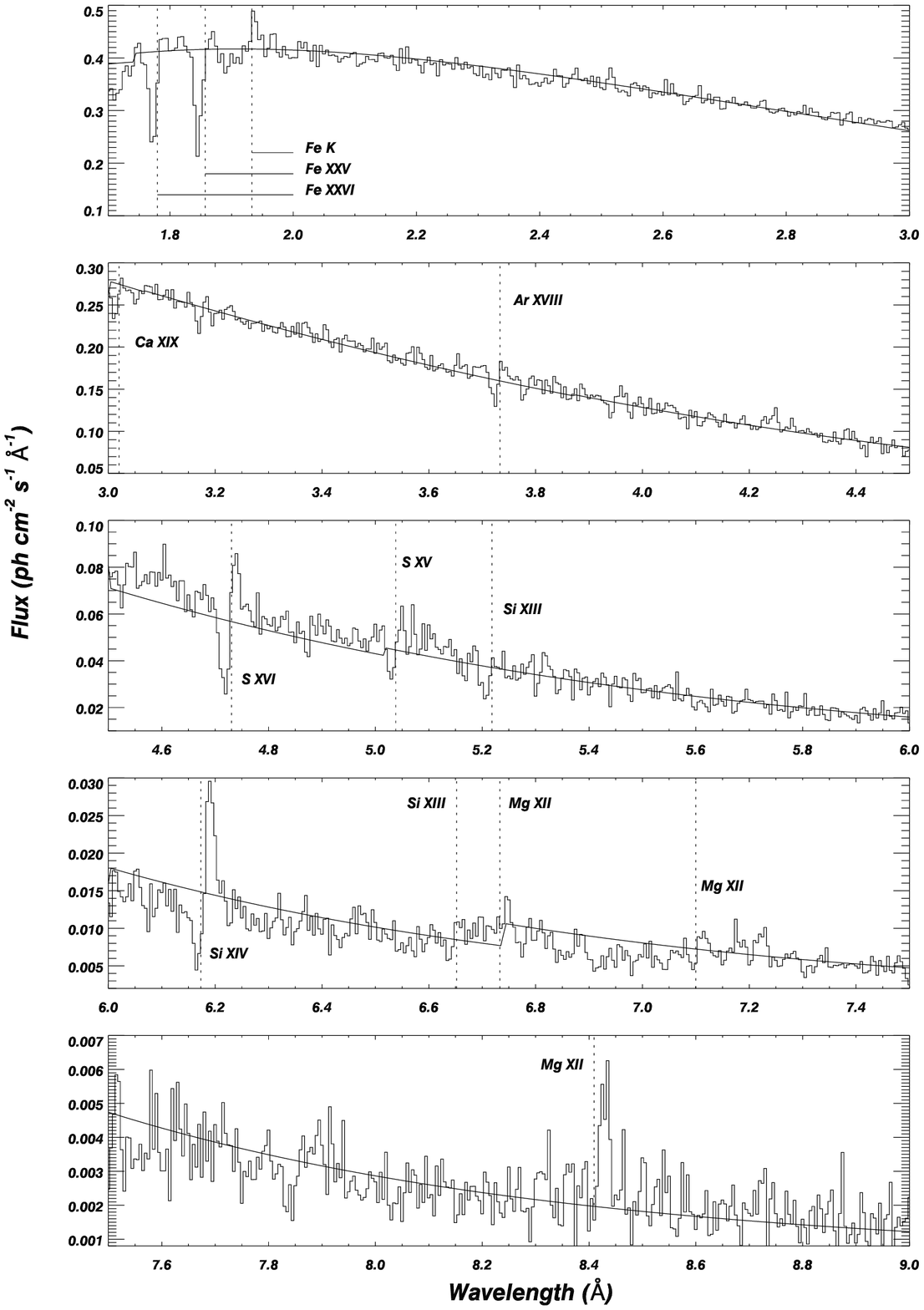}}
\vspace{9mm}
\figcaption{The co-added HEG 1st order spectrum from observation~II up to 9~\AA\ with all
identified spectral features marked. The two-blackbody model described in \S2.3 is also 
plotted, after we applied an empirical correction for pile-up (see \S2.2). 
\label{spectrum}}
\end{figure*}

\section{P~Cygni Line Analysis}

\subsection{Basic line measurements}

For the line analysis we binned all spectra into 0.005~\AA\ bins (which oversamples the HEG
resolution by a factor of two). All the analysis in this section has been performed on the 
raw data. The spectra are integrated over the entire exposure. We use lines detected
in the HEG 1st and MEG 3rd orders, except in observation~I where we use the 
MEG 1st order above 10~\AA. We detect a number of resonance line features in both 
observations, of which the brightest clearly show P~Cygni profiles (Paper~I). Figure~3 
shows the spectrum of observation~II from 1.7--9~\AA\ with all identified features 
marked. Table~1 includes all line features detected with a signal-to-noise ratio larger 
than three for both observations. Most of the lines are from H-like and He-like
species, although we do detect some Li-like and B-like species from elements with
high atomic numbers. For some of the lines the P~Cygni shape is less 
pronounced, and we find them either mostly in emission or absorption (see below).
Notable in observation~II is the detection of the forbidden line of He-like Mg~XI, 
which cannot appear as a P~Cygni line since it is not a dipole transition.   

\begin{table*}
\footnotesize
\begin{center}
{\sc TABLE~1\\
LINE IDENTIFICATIONS AND PROPERTIES}
\vskip 4pt
\begin{tabular}{llccccccc}
\hline
\hline
{} &
{} &
{Predicted $\lambda$} &
{Measured $\lambda$} &
{$v$} &
{$W_{\rm pc}$} &
{$f_{\rm a}$} &
{$N_{\rm abs}$} &
{$F_{\rm el}$} \\
{Ion} &
{Transition$^{\rm a}$} &
{(\AA)} &
{(\AA)} &
{(km s$^{-1}$)} &
{(m\AA)} &
{} &
{($\times 10^{16}$~cm$^{-2}$)} &
{($\times 10^{-4}$~ph~cm$^{-2}$~s$^{-1}$)} \\
{(1)} &
{(2)} &
{(3)} &
{(4)} &
{(5)} &
{(6)} &
{(7)} &
{(8)} &
{(9)} \\
\hline
\multicolumn{9}{c}{Observation~I} \\
Fe~XXVI  & H-like (Ly$\alpha$) &  1.78 & $1.7792\pm0.0013$ & $1010\pm340$ & $3.93\pm0.23$ & 0.25 & 8.57 & $15.53\pm1.25$ \\
Fe~XXV   & He-like             &  1.85 & $1.8543\pm0.0014$ & $1940\pm320$ & $6.27\pm0.31$ & 0.48 & 12.60 & $21.54\pm1.75$ \\
Ar~XVIII & H-like (Ly$\alpha$) &  3.73 & $3.7291\pm0.0021$ & $ 320\pm160$ & $3.26\pm0.25$ & 0.46 &  3.96 & $11.64\pm1.42$\\
S XV     & He-like             &  4.31 & $4.3526\pm0.0024$ & $1860\pm140$ & $13.53\pm0.37^{\rm b}$ & $\cdots$ & $\cdots$ & $\cdots$ \\  
S XVI    & H-like (Ly$\alpha$) &  4.73 & $4.7252\pm0.0026$ & $ 820\pm130$ & $4.51\pm0.34$ & 0.44 &  3.36 & $16.84\pm2.00$\\
Si~XIV   & H-like (Ly$\delta$) &  4.85 & $4.8571\pm0.0026$ & $  60\pm120$ & $1.70\pm0.22$ & $\cdots$ & $\cdots$ & $8.11\pm1.28$ \\
Si~XIV   & H-like (Ly$\gamma$) &  4.95 & $4.9467\pm0.0027$ & $ 240\pm120$ & $1.40\pm0.21$ & $\cdots$ & $\cdots$ & $6.02\pm1.23$ \\
S XV     & He-like (r)         &  5.04 & $5.0333\pm0.0027$ & $ 300\pm120$ & $2.81\pm0.24$ & 0.64 &  1.41 & $6.52\pm1.37$  \\
Si~XIV   & H-like (Ly$\beta$)  &  5.22 & $5.2080\pm0.0028$ & $ 340\pm110$ & $2.21\pm0.26$ & 0.68 &  5.25 & $4.10\pm1.32$ \\
Si~XIV   & H-like (Ly$\alpha$) &  6.18 & $6.1816\pm0.0032$ & $ 480\pm180$ & $12.03\pm0.84^{\rm c}$ & 0.42 & 3.59 & $15.64\pm0.95$\\
Mg~XII   & H-like (Ly$\gamma$) &  6.74 & $6.7338\pm0.0035$ & $ 760\pm 90$ & $3.49\pm0.30^{\rm c}$ & 0.46 & 13.77 & $9.94\pm0.67$ \\
Mg~XII   & H-like (Ly$\beta$)  &  7.10 & $7.0902\pm0.0038$ &    $\cdots$  &  $1.12\pm0.44$ & $\cdots$ &  $\cdots$ & $5.48\pm0.58$ \\
Fe~XXIV  & Li-like             &  7.36 & $7.3613\pm0.0037$ &    $\cdots$  &  $1.04\pm0.41$ & $\cdots$ &  $\cdots$ & $4.81\pm0.57$ \\
Ni XXIV  & B-like              &  7.55 & $7.5623\pm0.0038$ &    $\cdots$  &  $1.05\pm0.41$ & $\cdots$ &  $\cdots$ & $4.30\pm0.55$ \\
Mg~XII   & H-like (Ly$\alpha$) &  8.42 & $8.4080\pm0.0043$ & $ 710\pm 70$ & $9.78\pm0.75$ & 0.49 & 1.83 & $4.80\pm0.62$ \\
Mg~XI    & He-like (i)         &  9.23 & $9.2139\pm0.0047$ & $ 460\pm 70$ & $7.01\pm1.36$ & 0.46 &  $\cdots$  & $1.71\pm0.53$\\
Ne~X     & H-like (Ly$\delta$) &  9.48 & $9.4891\pm0.0048$ & $ 220\pm 60$ & $7.02\pm1.24$ & 0.43 & 27.12 & $1.85\pm0.49$ \\
Fe~XXIV  & Li-like             & 10.63 & $10.6341\pm0.0098$ & $1130\pm170$ & $6.77\pm1.01^{\rm d}$ & 0.49 & 0.85 & $1.45\pm0.45$ \\
Ne~X     & H-like (Ly$\alpha$) & 12.13 & $12.1082\pm0.0102$ & $740\pm 70$ & $11.23\pm1.25^{\rm d}$ & 0.44 & 0.91 & $2.52\pm0.48$ \\
\hline
\multicolumn{9}{c}{Observation~II} \\
Fe~XXVI  & H-like (Ly$\alpha$) &  1.78 & $1.7796\pm0.0013$ & $1010\pm340$ & $6.39\pm0.25$ & 0.85 & 36.23 & $3.35\pm1.15$\\
Fe~XXV   & He-like             &  1.85 & $1.8570\pm0.0014$ & $1940\pm320$ & $6.89\pm0.29$ & 0.76 & 21.88 & $5.73\pm1.76$\\
Ca~XIX   & He-like (r)         &  3.04 & $3.0201\pm0.0018$ & $ 890\pm200$ & $3.13\pm0.30$ & 0.72 &  6.61 & $2.22\pm0.66$\\
Ar~XVIII & H-like (Ly$\alpha$) &  3.73 & $3.7333\pm0.0021$ & $ 640\pm160$ & $5.56\pm0.42$ & 0.59 &  6.40 & $3.40\pm0.54$\\
S XVI    & H-like (Ly$\alpha$) &  4.73 & $4.7304\pm0.0026$ & $ 760\pm130$ & $25.07\pm1.13$ & 0.58 & 17.65 & $6.15\pm0.57$\\
S XV     & He-like (r)         &  5.04 & $5.0381\pm0.0027$ & $ 600\pm120$ & $7.42\pm1.33$ & 0.47 &  2.02 & $1.41\pm0.41$\\
Si~XIV   & H-like (Ly$\beta$)  &  5.22 & $5.2182\pm0.0028$ & $ 800\pm110$ & $10.63\pm1.22$ & 0.91 & 26.04 & $0.37\pm0.38$ \\
Si~XIV   & H-like (Ly$\alpha$) &  6.18 & $6.1783\pm0.0032$ & $ 440\pm100$ & $49.63\pm1.77$ & 0.36 & 12.71 & $3.14\pm0.15$ \\ 
Si~XIII  & He-like (r)         &  6.65 & $6.6522\pm0.0035$ & $ 630\pm 90$ & $14.79\pm2.09$ & 0.64 &  3.19 & $0.41\pm0.14$\\ 
Mg~XII   & H-like (Ly$\gamma$) &  6.74 & $6.7330\pm0.0035$ &      $\cdots$      & $8.27\pm1.49$ & $\cdots$ & $\cdots$ & $0.48\pm0.10$ \\
Mg~XII   & H-like (Ly$\beta$)  &  7.10 & $7.1000\pm0.0037$ & $ 200\pm 80$ & $16.04\pm3.09$ & 0.44 & 20.02 & $0.40\pm0.12$ \\
Mg~XII   & H-like (Ly$\alpha$) &  8.42 & $8.4094\pm0.0043$ & $ 710\pm 70$ & $35.59\pm4.87$ & 0.32 &  4.36  & $0.40\pm0.09$ \\
Mg~XI    & He-like (r)         &  9.17 & $9.1700\pm0.0047$ &      $\cdots$      & $12.03\pm4.50$ & $\cdots$ &  $\cdots$ & $0.15\pm0.06$ \\
Mg~XI    & He-like (i)         &  9.23 & $9.2319\pm0.0047$ &      $\cdots$      & $14.29\pm4.35$ & $\cdots$ &  $\cdots$ & $0.21\pm0.08$ \\
Mg~XI    & He-like (f)         &  9.31 & $9.3088\pm0.0048$ &      $\cdots$      & $14.79\pm5.99$ & $\cdots$ &  $\cdots$ & $0.23\pm0.08$\\
Ne~X     & H-like (Ly$\delta$) &  9.48 & $9.4861\pm0.0048$ &      $\cdots$      & $17.29\pm4.76$ & $\cdots$ &  $\cdots$ & $0.14\pm0.06$\\
\hline
\vspace*{0.02in}
\end{tabular}
\parbox{5.2in}{
\small\baselineskip 9pt
\footnotesize
\indent
$\rm ^a${From Mewe (1994).}\\
$\rm ^b${Affected by a chip gap, and results are unreliable.}\\
$\rm ^c${Added from MEG 3rd order.}\\
$\rm ^d${Added from MEG 1st order.}\\
}
\end{center}
\setcounter{table}{1}
\normalsize
\centerline{}
\end{table*}

Table~1 lists each ion in column~1, the transition name in column~2
(``H-like'' denotes ions with only one electron left, ``He-like'' with two, and so on), 
and the predicted wavelength from the SPEX (Mewe 1994) line list in column~3. In 
column~4 we list the measured wavelength, which represents the intersection between the 
emission and the absorption with the continuum. The intersection
with the continuum is generally a good estimate of the rest-wavelength position. This is underlined
by the fact that on average the profiles appear symmetric. In order to quantify
the properties of the P~Cygni lines, we normalized each profile by dividing the data by the underlying local
continuum (where we simply applied a polynomial fit) and then fitted the absorption and emission components
with Gaussians. Within the statistical uncertainties, simple Gaussian
functions proved to be sufficient for the fits. The velocities listed in column~5 represent 
the difference between the measured wavelength and the peak position of the Gaussian fitting 
the absorption part of the line (note, of course, that the maximum velocity to which absorption 
is observed is higher than this); lines without clear P~Cygni shapes do not have velocity 
values listed. Most emission peaks appear slightly above the 
expected wavelength in column~3; in a few cases they match well, and
here we have probably slightly underestimated the velocity.
We find velocities of 200--1900~km~s$^{-1}$.

We determined the equivalent widths of the lines from the fits. They are listed in
column~6 of Table~1. The values are measured from HEG 1st order data unless noted otherwise.
The equivalent width of a P~Cygni line, $W_{\rm pc}$, is defined as the sum of the equivalent widths of the emission and the
absorption (taking both to be positive quantities). For the 
analysis below we also list the contribution of absorption as a fraction of $W_{\rm pc}$
($f_{\rm a}$ in column~7 of Table~1).  
The Fe-line area below 2~\AA\ is treated separately below. 

\subsection{High-flux versus low-flux states}

In this section we analyze the variations of the P~Cygni lines between observation~I and observation~II.
The measured wavelengths (as well as the ones for the peak and valley positions) of the same transitions 
in both observations agree to within the stated uncertainties. 
The most dramatic change is observed in the equivalent widths
of the lines. In Figure~4 we compare the P~Cygni profiles for the lines with the highest equivalent
widths in both observations, which are the Lyman~$\alpha$ lines of S~XVI (left), Si~XIV (middle), and 
Mg~XII (right). Pile-up effects for S~XVI and Si~XIV cannot account for the strong variability 
of the lines. The equivalent widths in observation~I are generally lower than 
in observation~II throughout the whole bandpass. The equivalent widths of the 
three Lyman~$\alpha$ lines in Figure~4 differ by a factor of $\approx 4$, whereas 
the other lines typically differ by a factor of $\approx 2$.
 
\subsection{Emission lines}

The fluxes of the emission lines are shown in column~9 of Table~1. The lines in observation~I 
appear bright with fluxes from $1.5\times10^{-4}$~ph~cm$^{-2}$~s$^{-1}$ at 10.63~\AA\ to 
$2.2\times10^{-3}$~ph~cm$^{-2}$~s$^{-1}$ at 1.85~\AA. The general trend of declining line 
flux with increasing wavelength is largely due to absorption by the interstellar column 
density. The brightest lines appear to be the Lyman~$\alpha$ lines of Fe~XXVI, S~XVI, 
and Si~XIV, as well as He-like Fe~XXV. In observation~II the lines' fluxes dropped 
significantly to a range from $1.5\times 10^{-5}$~ph~cm$^{-2}$~s$^{-1}$ to 
$6.2\times 10^{-4}$~ph~cm$^{-2}$~s$^{-1}$. There is a trend for the largest decline in 
line flux to happen toward longer wavelengths; above 6.5~\AA\ the decline is 
$\approx$~12--20 while below 5~\AA\ the decline is $\approx$~3--5. 

\centerline{\epsfxsize=8.5cm\epsfbox{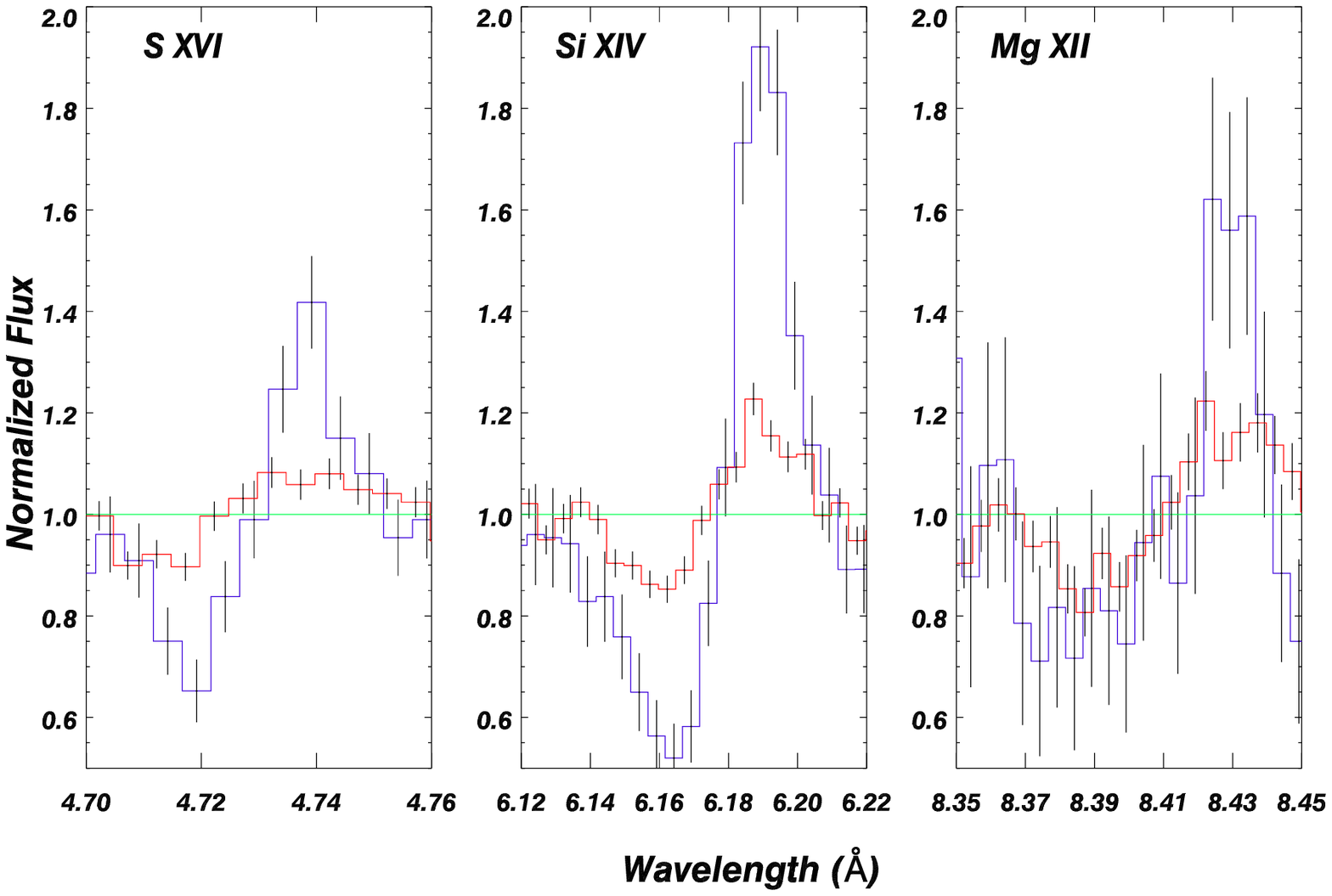}}
\figcaption{Relative P~Cygni line strengths of the Lyman~$\alpha$ transitions in S~XVI, S~XIV, and Mg~XII for
observation~I (red) and observation~II (blue).
\label{pcygni}}
\vspace{3mm}

Of particular interest is the He-like Mg~XI triplet, because it is the only one we can 
clearly resolve and appears free of contaminating features; for comparison, the Si~XIII 
triplet is confused with the Mg~XII Lyman~$\gamma$ line as well as the Si~K edge. He-like 
triplets are sensitive to temperature and density and thus can be used as plasma 
diagnostics; we use the calculations by Porquet \& Dubau (2000). Calculation of the 
$G=(i+f)/r$ ratio ($i$ is the flux of the intercombination line, $f$ the flux of the 
forbidden line, and $r$ the flux of the recombination line) and $R=i/f$ ratio allows 
estimation of the temperature and density, respectively (Gabriel \& Jordan 1969). 
Our best constraints come from observation~II since during this observation
we detect all the lines in the triplet as shown in Figure~5. We compute a $G$ ratio 
of $2.9\pm 1.0$ and an $R$ ratio of $1.1\pm 0.34$, implying a temperature of 
$(3.8\pm1.1)\times 10^6$~K and a density of $(1.9\pm0.9)\times 10^{13}$~cm$^{-3}$.
A $G$ ratio of $\approx 3$ indicates a photoionized plasma dominated by recombination 
(Bautista \& Kallman 2000) and justifies the use of the Porquet \& Dubau (2000) 
calculations, which are specifically designed for a recombining plasma.
During observation~I we only detect the intercombination line at 9.23~\AA\ and use 
the $1\sigma$ uncertainty in the continuum as an upper limit for the detection of 
the forbidden and resonance lines. We estimate an $R$ ratio of $>0.5$, implying 
a density of $<4\times 10^{13}$~cm$^{-3}$. The $G$ ratio is not well constrained
during observation~I. 

Figure~5 shows that the Mg~XI triplet probably does not suffer from blending, which 
would mainly stem from possible lower-shell absorption events. These are unlikely 
given the high ionization states observed. Fe~L emission lines are unlikely 
to appear as blends, because Kallman et~al. (1996) predicted that recombination 
emission is dominated by K-shell ions and not L-shell ions. We cannot exclude 
the possibility that resonance scattering affects the line ratios; this would
tend to increase the flux of the resonance line and thus lower the $G$ ratio. 
The consequence would be that the temperature is higher or, in the extreme 
case, that the Porquet \& Dubau (2000) calculations are not applicable because 
the plasma is not predominately recombining. However, the resonance line does 
not appear very strong in the spectrum, and a correction of $G$ due to resonance 
scattering seems unlikely to be highly significant.  

\centerline{\epsfxsize=8.5cm\epsfbox{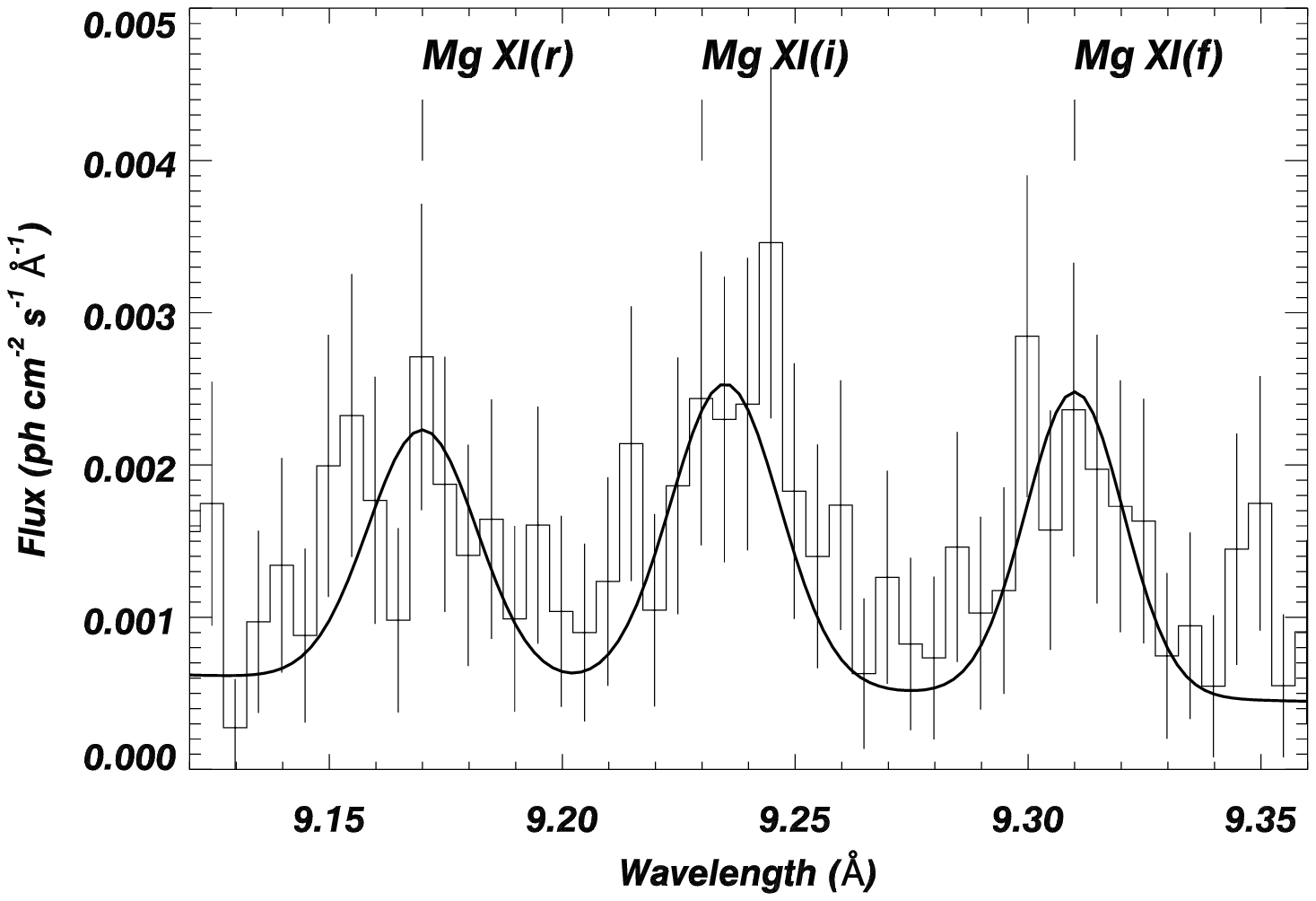}}
\figcaption{The He-like Mg~XI triplet in observation~II with resonance (r), intercombination (i), and
forbidden (f) lines.
\label{mgxi}}
 
\subsection{Absorption column densities}

The measurements of the absorption-line equivalent widths allow estimation of the 
ionic column densities in the absorbing medium. For an absorption line on the 
linear part of the curve of growth, the relation between equivalent width and 
column density is

\begin{equation}
 {W_{\lambda}\over \lambda} = {{\pi e^2}\over{m_{\rm e}c^2}} N_{\rm j} \lambda f_{\rm ij} = 8.85\times10^{-13}N_{\rm j} \lambda f_{\rm ij}
\end{equation}

\noindent
where $N_{\rm j}$ is the column density of ion species $j$, $f_{\rm ij}$ is the oscillator strength 
of the transition, and $W_{\lambda}$ is the equivalent width of the absorption 
line (e.g., Spitzer 1978; see also Lee et~al. 2001). In Table~1 we list the column density, $N_{\rm abs}$, for 
each ion species derived using this formula. 
 
Several effects may lead to systematic underestimates of column densities
when the formula above is used. Given the observed P~Cygni profiles, for example, it
is likely that line emission fills in some of the absorption; it is difficult to 
correct for the resulting column-density underestimates precisely. Furthermore, 
line saturation can also lead to an underestimate of the column density. The 
observed lines could be saturated even though they do not appear ``black'' because 
of multiple light paths to the observer (e.g., the electron scattering of some 
X-rays around the absorbing material; see \S1). The observed lines 
might also have unresolved substructure composed of narrower, saturated lines. In 
fact, since we observe lines from high-order transitions in the Lyman series, the 
Lyman~$\alpha$ lines are probably saturated; the oscillator strengths decrease by 
a factor of $\approx 30$ from the 1s-2p (Lyman~$\alpha$) to the 1s-4p (Lyman~$\delta$)
transitions. We therefore see the column densities in Table~1 more as lower limits, 
and we note that the column densities derived from the high-order Lyman transitions 
probably have the least systematic error. For the determination of column densities
we thus rely only on the highest-order Lyman transition detected.

In order to estimate equivalent hydrogen column densities, we must correct for the 
ionization fraction of a given ion. From the analysis of the Mg~XI triplet, it is 
likely that we observe a photoionized plasma. In the following we use the calculations 
by Kallman \& Bautista (2001) for such a plasma which provide a representative 
ionization structure. For a temperature of $3.8\times 10^{6}$~K (see \S3.3), 
we find an ionization parameter of $\xi=L/nR^2\approx 20,000$~erg~cm~s$^{-1}$, which is 
consistent with the fact that we observe mostly H-like ions. We find a similar value 
of $\xi$ when we compare our abundance ratios 
[AR=$N_{\rm abs}$(H-like)/$N_{\rm abs}$(He-like)] with their calculations. 
Table~2 shows the ionization fractions for Mg, Si, S, and Fe and the derived equivalent 
hydrogen column densities as determined from the observed abundance ratios; 
the column densities are given for both a solar-abundance 
distribution from Morrison \& McCammon (1983) and an interstellar-abundance 
distribution from Wilms, Allen, \& McCray (2000).

The equivalent hydrogen column densities in Table~2 appear to increase significantly 
between observation~I and observation~II. However, as discussed in \S3.3, we only
have limited constraints on the ionization structure of the gas during observation~I
because we do not detect all the lines in the Mg~XI triplet. We cannot rule out, 
for example, a reduction in the ionization fraction during observation~II that
explains all the observed changes. Some of the derived column densities should 
be taken with caution. It is unusual, for example, that we observe a strong 
Mg~XII 1s-3p transition, whereas the 1s-2p transition is missing. In fact if 
we discard the result for this line, the derived column density using Mg would 
be similar to those computed using the other elements.

\vbox{
\footnotesize
\begin{center}
{\sc TABLE~2\\
IONIZATION FRACTIONS AND EQUIVALENT HYDROGEN COLUMN DENSITIES}
\vskip 4pt
\begin{tabular}{llccccc}
\hline
\hline
{} &
{} &
{} &
{$N_{\rm H}^{\rm c}$} &
{} &
{} &
{$N_{\rm H}^{\rm c}$} \\
{Ion} &
{AR$^{\rm a}$} &
{$f_{\rm ion}^{\rm b}$} &
{($10^{22}$~cm$^{-2}$)} &
{AR$^{\rm a}$} &
{$f_{\rm ion}^{\rm b}$} &
{($10^{22}$~cm$^{-2}$)} \\
\hline
\multicolumn{1}{c}{} & \multicolumn{3}{c}{Observation~I} & \multicolumn{3}{c}{Observation~II} \\
Fe~XXVI  &  0.68 & 0.327  & 1.5 (1.9) & 1.7 & 0.491 & 4.6 (5.6) \\
Fe~XXV   &   $\cdots$ & 0.514  & $\cdots$ &  $\cdots$ & 0.277 &  $\cdots$ \\
S XVI    &  2.4 & 0.480  & 0.7 (1.1) & 8.7 & 0.290 & 6.4 (9.9) \\
S XV     &   $\cdots$ & 0.199  & $\cdots$ &  $\cdots$ & 0.032 &  $\cdots$ \\
Si~XIV   &  5.1$^{\rm d}$ & 0.310  & 0.7 (1.4) &  8.2& 0.244 & 7.9 (15.8) \\
Si~XIII  &   $\cdots$ & 0.035  & $\cdots$ &  $\cdots$ & 0.019 & $\cdots$ \\
Mg~XII   &  14.5$^{\rm d}$ & 0.237  & 3.0 (4.7) & 22.4$^{\rm d}$& 0.162 & 7.2 (11.5) \\
Mg~XI    &   $\cdots$ & 0.016  & $\cdots$ &  $\cdots$ & 0.007 & $\cdots$ \\
\hline
\vspace*{0.02in}
\end{tabular}
\parbox{3.2in}{
\small\baselineskip 9pt
\footnotesize
\indent
$\rm ^a${$N_{\rm abs}$(H-like)/$N_{\rm abs}$(He-like)}\\
$\rm ^b${Ionization fractions from Kallman \& Bautista (2001).}\\
$\rm ^c${Equivalent hydrogen column density for a solar-abundance distribution
(Morrison \& McCammon 1983) and an interstellar-abundance distribution
(Wilms, Allen, \& McCray 2000; in parentheses).}\\ 
$\rm ^d${Estimated lower limits.}\\
}
\end{center}
\setcounter{table}{3}
\normalsize
\centerline{}
}

\subsection{The Fe~K line region}

The Fe~K line region is especially interesting for spectroscopy because Fe has quite high
fluorescent yields (Auger destruction is not as dominant as in lower $Z$ elements) as 
well as ions that can survive at high ionization parameters. Figure~6 (top) shows the 
Fe~K region from 1.7--2.0~\AA\ as observed with the HEG during observation~I. 
The model (red curve) includes the continuum level as fitted over the entire available
bandpass (see \S2.3). In addition we applied an empirical correction to account for
pile-up (see \S2.2) which we computed by matching featureless continuum stretches in 
the 1.97--3.00~\AA\ range. Below 1.97~\AA\ additional Gaussians were fitted locally 
and then added to the model. At the short wavelength side of the Fe edge we additionally 
adjusted the model to match the continuum level. We observe an Fe~K line centered 
at 1.94~\AA, which is consistent with the charge states Fe~I--X (e.g., House 1969). 
The line has an equivalent width of 1.9~m\AA. The spectrum 
also shows P~Cygni lines of Fe~XXV and Fe~XXVI with the properties 
given in Table~1. We resolve a neutral Fe edge at 1.75~\AA\ with an optical depth 
of $\tau=0.035\pm0.017$, implying an equivalent hydrogen column density 
of $3.4\times 10^{22}$~cm$^{-2}$. After subtraction of the interstellar 
column density, the column densities of neutral and ionized gas appear 
similar. 

Figure~6 (bottom) shows the same spectral region for observation~II. The model was 
calculated in a similar way to that for observation~I; however, no correction for
pile-up was needed. We detect an Fe~K line with a similar 
position and equivalent width as that observed during observation~I.
However, we do not see obvious P~Cygni lines from Fe~XXV and Fe~XXVI but mostly 
resonance absorption with significantly increased equivalent width (see Table~1). The 
resonance-line positions are the same as observed during observation~I indicating little 
change in velocity. The fact that we do not see much emission from ionized Fe 
in observation~II may at first seem surprising, since we see P~Cygni lines in 
lower $Z$ ion species. However, if one compares the fluxes of the Lyman~$\alpha$ lines 
and He-like lines of S, Si, and Mg in Table~1, then the decrease in flux is consistent 
with that observed in these other ions. 

\vspace{10mm}
\centerline{\epsfxsize=8.5cm\epsfbox{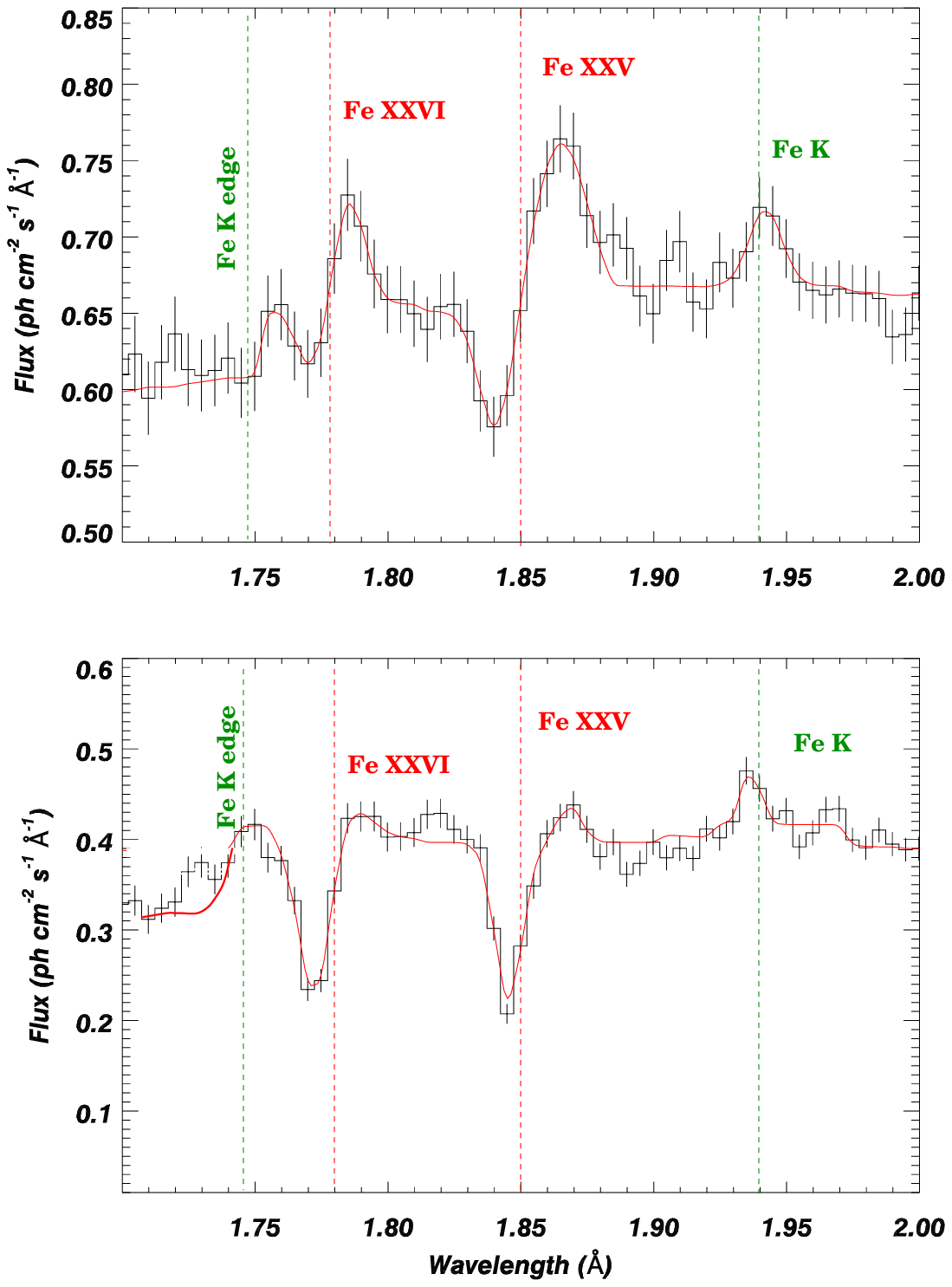}}
\figcaption{Spectral fits of the Fe-K line regions of observation~I (top) 
and observation~II (bottom) using the HEG data. 
\label{feI}}

A substantial increase in the depth of the neutral Fe~K edge is observed 
between observation~I and observation~II. During observation~II the depth is 
$\tau=0.108\pm 0.028$. To derive this value we had to take into account an 
unidentified feature at 1.73~\AA; because of this feature we had to force 
the edge to match the high-energy continuum. It is difficult 
to derive a column density from the edge depth. Due to the
partial-covering nature of the absorption and the fact that the neutral absorbing 
column density is expected to be $\approxgt 10^{24}$~cm$^{-2}$ during observation~II
(see \S4.4 of Shirey et~al. 1999b), the true column density is 
likely to be substantially larger than expected from simple consideration of
the apparent edge depth. A column density of $\approxgt 10^{24}$~cm$^{-2}$
will absorb much of the direct continuum at wavelengths longward of the edge, and 
due to this ``saturation'' effect the true depth of the edge (and hence the column 
density) can be severely underestimated. 

\section{Short-term Variability of the Lines}

We have also studied short-term variability of the P~Cygni lines on time scales 
within the 30~ks and 15~ks exposure times. To do this, we apply a ``sliding-window'' 
technique, where we define a fixed exposure-time window that is sufficiently long
to receive a significant signal from the bright lines. At a set of equally spaced
times during the observation (separated by a constant ``time step''), we then create 
spectra using the data within the exposure-time window (the exposure-time window
is centered on each of the equally spaced times). In the case of observation~I, we 
selected an exposure-time window length of 5~ks and a time step of 1~ks (note that 
the resulting spectra have some overlap). Within a total exposure time of 30~ks, 
this allows us to create 24 spectra. Using these spectra, we then created
three-dimensional ``spectral maps'' where the $x$-axis is the wavelength of 
each spectrum in 0.015~\AA\ steps, the $y$-axis is the time in 1.0~ks 
steps, and the $z$-axis is the photon flux in each pixel. 

Figure~7 shows spectral maps for three line regions: the Fe line region (left),
the Si~XIV Lyman~$\alpha$ line region (middle), and the Mg~XII Lyman~$\alpha$ line region (right).  
The colors represent the flux levels from low (green to black) to high (orange to yellow) on a 
logarithmic scale. The scale ranges from $\approx$~0.1--0.8~ph~cm$^{-2}$~s$^{-1}$.
The color changes in these maps in the vertical direction show the variability 
of the spectra. This variability appears somewhat smoothed since the windows overlap. 

The spectral continuum is clearly changing; this is indicated by the large-scale color
distribution in each map. At short wavelengths (left map), the continuum stays high throughout
most of the observation and then drops rapidly in the last 5~ks. At intermediate
wavelengths (middle map), the continuum remains low throughout the first one-third of the 
observation, increases rapidly around the middle of the observation, and remains
high throughout most of the rest of the observation. At longer wavelengths (right map), this
trend is continued; the continuum stays low throughout the first half of the observation.
This qualitative assessment of the continuum variability is consistent with the 
\xte results of Shirey et~al. (1999a) who find the spectra moving along 
spectral branches in the X-ray color-color diagram.   

The three maps in Figure~7 show that the shapes of the P~Cygni lines are 
highly variable on time scales of 2--3~ks. In the time maps at the
top of Figure 7 we observe a correlation 
of the emission and an anti-correlation of the resonance absorption with the 
local continuum flux. When the local continuum is low, the emission-line 
parts of the P~Cygni profiles are weak; they grow stronger with increasing continuum flux. 
The absorption-line parts show the opposite behavior; their amplitudes are larger at lower 
continuum fluxes, and absorption practically vanishes at high flux levels. This behavior 
on short time scales reflects the same general pattern we observed by comparing the 
high-flux state in observation~I with the low-flux state
in observation~II. This also explains why for Fe~XXV the emission line reaches its peak
flux about 7~ks {\it before\/} we observe the peak in the Si~XIV emission line. The line 
flux is directly connected with the local continuum flux, and in this case the spectral 
change pattern creates this time difference.     

Figure~7 also shows the spectra used to make the spectral maps. The time sequence of 
the Si~XIV line at 6.18~\AA\ shows that the absorption part of the profile has 
vanished by about 12~ks into observation~I, which would imply a column density 
change of Si~XIV from about twice the amount in Table~1 to nearly zero on a time scale 
of $\approx 3$~hr. The fact that the flux of the local continuum is anti-correlated 
with the absorption amplitude suggests that changes in the ionization fraction, 
rather than changes in column density, are primarily responsible for the changes 
in absorption strength (although we cannot strictly rule out the latter). 

\begin{figure*}
\centerline{\epsfxsize=17.5cm\epsfbox{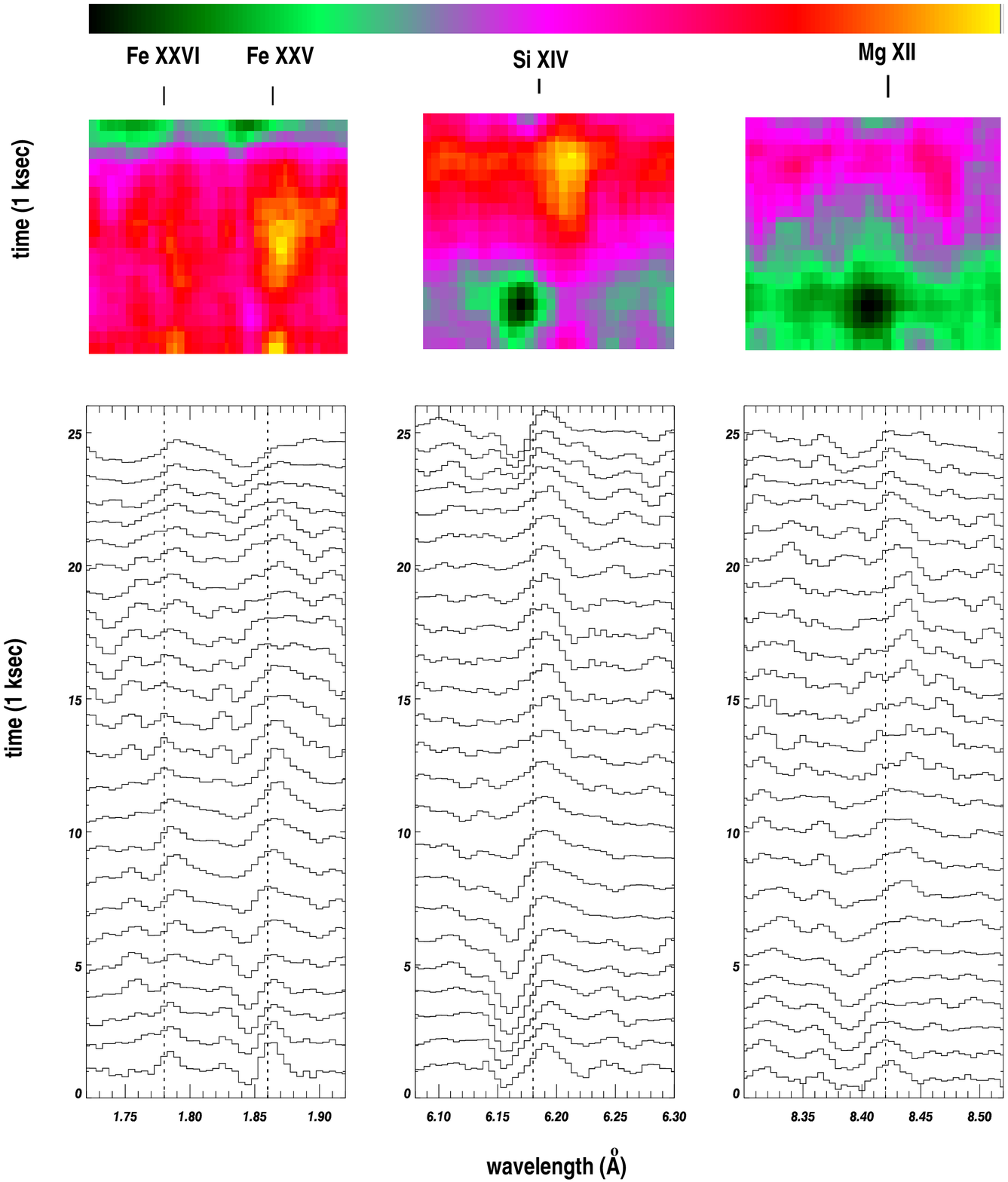}}
\figcaption{Top: Spectral maps of selected regions of observation~I: the Fe-line 
region (left), the Si~XIV region (middle), and the Mg~XII region (right). Bottom: The spectral
profiles from each of the spectral maps above. The time step between
each profile is 1~ks.  
\label{timemap}}
\end{figure*}

The variability of the Si~XIV line during observation~II is shown in Figure~8. 
Although during observation~II the overall X-ray flux is substantially 
lower than during observation~I, the ionization parameter appears to be varying 
over approximately the same range. The emission-line fluxes during observation~II, 
on the other hand, are lower by significant factors (see \S3.3). 

Figure~8 also illustrates the correlation of spectral hardness 
and absorption strength for observation~II. We computed a hardness
ratio by dividing the flux in the 1.7--3~\AA\ band by that in the 3--8~\AA\ band; this 
ratio is similar to the soft-band hardness ratio computed by Shirey et~al. (1999b). 
The flux is computed in the 1.7--8~\AA\ band.
We used the same sliding-window technique as used for the spectral maps, and 
thus in Figure~8 the sequence of data points in the hardness-ratio diagram 
can be directly matched to the sequence of line profiles. The data points in 
Figure~8 are connected by dotted lines indicating the time sequence. In the 
inset of Figure~8 we also plot the softest (spectrum~1) and hardest (spectrum~2) 
spectra during observation~II, which occurred near the beginning and the end 
of the observation, respectively. Figure~8 shows that while the flux is 
generally decreasing throughout the observation, the hardness ratio increases 
until it reaches a plateau at $\approx 2.3$. This spectral behavior is consistent 
with that studied during dip periods with \xte (Shirey et~al. 1999b). 
The more the continuum hardens, the more the absorption in the P~Cygni profiles 
decreases. 

During observation~I we see similar variations of the P~Cygni line on similar 
time scales. However, here the spectrum is dominated by the soft component, and 
spectral variations of the hard component cannot be directly measured.

\begin{figure*}
\centerline{\epsfxsize=16.5cm\epsfbox{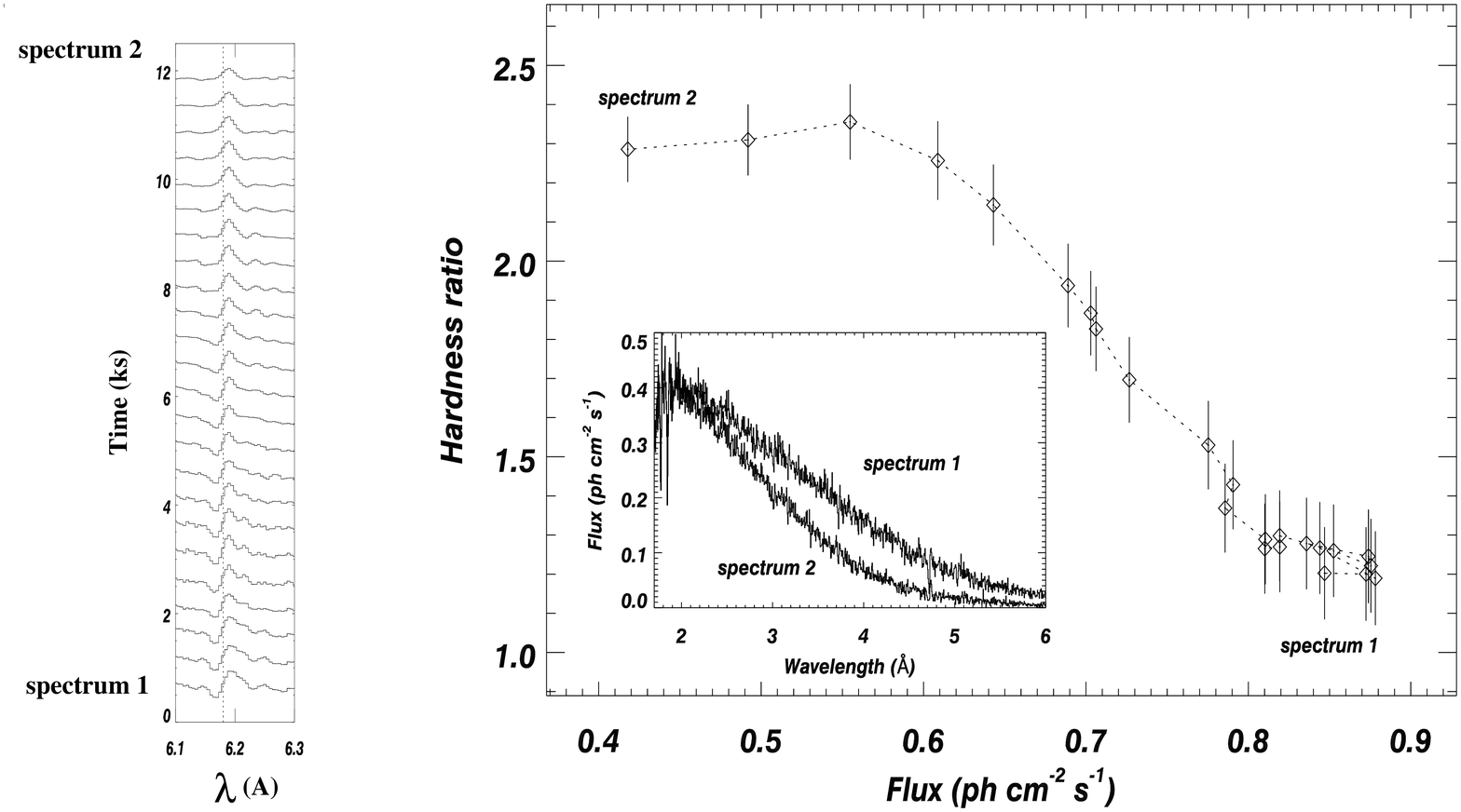}}
\figcaption{Si~XIV line profiles and hardness ratio versus intensity plot 
for observation~II. The flux is computed between 1.7 and 8~\AA.
We used the same sliding-window technique as used for 
Figure~7. Each data point in the right panel can be directly matched to one 
of the line profiles in the left panel. The dotted lines connect the data 
points in order of increasing time. The inset compares the the first (spectrum~1) 
and the last (spectrum~2) spectra in the sequence.
\label{pvariation}}
\end{figure*}

\section{Discussion}

\subsection{Refined constraints on the accretion-disk wind model}

In Paper~I we argued that the P~Cygni profiles are most likely produced by a
high-velocity outflow from an accretion disk viewed in a relatively edge-on
manner. The central X-ray source illuminates the disk and produces a wind
driven by both thermal and radiation pressure (e.g., Begelman et~al. 1983). 
The intermediate-temperature ($\sim 5\times 10^6$~K) part of this wind 
produces the H-like and He-like X-ray lines (e.g., Raymond 1993; 
Ko \& Kallman 1994) with P~Cygni profiles. Our current data remain 
generally consistent with this scenario, although we recognize that 
the strong wind variability seen from Cir~X-1 is not usually accounted for
in the predominantly time-independent theoretical studies; we use these 
studies as general guides to interpretation rather than for strict 
comparison purposes. 

Our new analyses, combined with other recent work on accretion-disk winds
and photoionized plasmas, have allowed us to extend some of the findings
in Paper~I. For example, one of the 
potential problems with the disk-wind model noted in \S3 of Paper~I
can now be ameliorated. There we noted that likely values for the wind's
launching radius ($r_{\rm launch}\approx 10^5$~km),
launching density ($n_{\rm launch}\approxgt 10^{15}$~cm$^{-3}$), and
ionization parameter ($\xi\lesssim 1000$~erg~cm~s$^{-1}$)
led to a wind that was optically thick to electron scattering; most line
photons attempting to traverse the wind would then be Compton-scattered
out of the line. However, our analysis in \S3.4, which uses the
calculations of Kallman \& Bautista (2001) rather than those of
Kallman \& McCray (1982), now suggests that a significantly higher
($\xi\approx 20,000$~erg~cm~s$^{-1}$) ionization parameter is likely for
much of the gas creating the P~Cygni lines. The required column density
along the line of sight is thus correspondingly reduced. Hydrogen
column densities of $\approx 10^{23}$~cm$^{-2}$ can now be accomodated
with only moderate shielding of the wind from the full X-ray continuum
or clumping of the wind.

The density measurement in the line-emission region of
$\approx~2\times 10^{13}$~cm$^{-3}$ using the Mg~XI triplet (see \S3.3)
provides a further independent constraint on the disk-wind model.
Following the arguments in \S3 of Paper~I with the new ionization parameter
of $\xi\approx 20,000$~erg~cm~s$^{-1}$, the expected launching density is
$n_{\rm launch}\approxgt 5\times 10^{13}$~cm$^{-3}$. 
The measured density in the
line-emission region is somewhat, but not greatly, lower than the
launching density. Note that since the launching density depends
upon the assumed launching radius, the general consistency found
above supports our adopted $r_{\rm launch}\approx 10^{5}$~km (chosen
to make the observed terminal velocity of the wind comparable to the
escape velocity at $r_{\rm launch}$; see \S3 of Paper~I).

Gratings observations of other neutron star LMXBs thus far have not generally 
revealed P~Cygni lines as prominent as those observed from Cir~X-1
(e.g., Cottam et~al. 2001; Marshall et~al. 2001; Paerels et~al. 2001;
Schulz et~al. 2001; Sidoli et~al. 2001). It 
is difficult to account for this difference quantitatively, but it may 
be due to a combination of mass accretion rate (relative to the Eddington 
rate) and inclination. The high, and perhaps super-Eddington, mass 
accretion rate of Cir~X-1 allows an unusually powerful outflow to be 
driven from the accretion disk, and a relatively edge-on geometry is 
optimal for the viewing of an equatorial outflow. 

\subsection{Line and continuum variability}

Given the results from \asca and \xte (Brandt et~al. 1996; Shirey et~al. 1999b), 
the large difference in overall X-ray flux between observation~I and
observation~II is likely to be due mostly to a change in the amount of 
absorbing material along the line of sight. As discussed by several 
authors (see \S1), this material is probably associated with a thickening 
of the accretion disk induced by the strong mass transfer occurring at zero 
phase. The change in Fe~K edge depth observed by \chandra is consistent with 
a large change in the amount of absorbing material, although it is
difficult to constrain the exact column density during observation~II. 
The fact that we observe the same basic P~Cygni lines during observation~I and 
observation~II supports the idea that the underlying ionizing continuum
source has not changed dramatically in strength. Furthermore, the fact that 
the line equivalent widths are significantly larger during observation~II 
than observation~I (see Figure~4) is generally consistent with the idea that 
blockage of direct continuum emission is occurring during observation~II. 
However, it is worth noting that the emission-line fluxes do drop significantly 
between observation~I and observation~II (see \S3.3). This can be understood 
without invoking continuum changes if the absorbing material screens ionizing 
photons from reaching a significant fraction of the line-emitting gas (this 
screening could be stronger at longer wavelengths, explaining the trend 
with wavelength noted in \S3.3). 

However, while it seems clear that changes in absorption play a critical
role in causing much of the variability of Cir~X-1, some changes in underlying 
continuum strength and shape are probable as well. These are particularly
likely during observation~I since it was made while Cir~X-1 was undergoing 
an X-ray flare above its ``base'' level (see Figure~1 of Paper~I). Our data 
are not ideal for probing the nature of the continuum variability in detail, 
but Shirey et~al. (1999a) have demonstrated that Cir~X-1 shows spectral 
variations that classify it as a ``Z'' type LMXB. The \chandra data strongly 
suggest that much of the short-term P~Cygni line variability is caused by spectral changes 
in the local continuum which lead to changes in the ionization level
of the wind (see \S4), although it is difficult to rule out entirely changes in the 
wind geometry as well. The observed short-term changes of the lines can be 
dramatic, causing them to vary from being almost completely in absorption 
to almost completely in emission. Modeling this variability in detail will 
be a challenge. 

\subsection{Comparisons with the spectral features seen from superluminal jet sources}

The Fe~K line region in the low-flux state spectrum of Cir~X-1 shows strong, blueshifted 
resonance absorption features from Fe~XXV and Fe~XXVI. Significant line emission during 
the low-flux state is only seen from neutral or nearly neutral Fe. Such spectral behavior 
is reminiscent of that seen from Galactic superluminal jet sources containing a 
black hole like GRS~1915+105 (Kotani et~al. 2000; Lee et~al. 2001) and GRO~J1655--40 
(Ueda et~al. 1998; Yamaoka et~al. 2001). Similar Fe absorption has also now 
been observed in GX~13+1 (Ueda et~al. 2001), a neutron star LMXB that has been 
classified as an atypical atoll source (Hasinger \& van der Klis 1989), and  
MXB~1659--298 (Sidoli et~al. 2001), also probably an atoll source; these results
show that these spectral features can appear independent of the nature of the compact 
object and the existence of jets. In most of these cases it has been argued that 
the absorption lines come from a highly ionized plasma layer extending to 
large heights above the equatorial plane (Yamaoka et~al. 2001). Specifically,
a geometrically thick accretion flow at $\sim 10^4$ Schwarzschild radii is assumed, 
which in the case of GRO~J1655--40 amounts to $\sim 10^5$ km. This is of the same 
magnitude as our launching radius for the wind in Cir~X-1, and it is worth 
considering the possibility that the material creating the Fe X-ray absorption 
lines in superluminal sources might generally be outflowing rather than inflowing. 
An equatorial wind scenario appears consistent with the gratings observations of 
GRS~1915+105 (Lee et~al. 2001); if the outflow velocity for GRS~1915+105 is
assumed to be the same as that for Cir~X-1, the outflow would need to be less
inclined relative to the line of sight to be consistent with the 770~km~s$^{-1}$ 
upper limit on the blueshift. The strong Fe~XXVI absorption feature from 
GRS~1915+105 shows variations on similar time scales to those observed from Cir~X-1. 
 
Superluminal jet sources containing black holes have not shown as many spectral
features as we observe from Cir~X-1; typically only features from highly 
ionized iron are seen. This could plausibly be explained if their winds are 
more highly ionized due to their larger luminosities and different continuum
shapes, so that iron is the only abundant element not fully stripped of its
electrons. 

\acknowledgments
We thank all the members of the \chandra\ team for their enormous efforts.
We gratefully acknowledge the financial support of CXC grant GO0-1041X (WNB, NSS) and
Smithsonian Astrophysical Observatory contract SV1-61010 for the CXC (NSS).

\end{document}